\documentstyle[aps]{revtex}
\include{psfig}
\begin{document}
\draft
\bibliographystyle{prsty}
\title{ Dynamic Phase Transition in a Time-Dependent Ginzburg-Landau
Model in an Oscillating Field}
\author{H.~Fujisaka,$^{1,*}$ H.~Tutu,$^{1,\dag}$ 
and P.~A.\ Rikvold$^{2,\ddag}$}
\address{
$^1$Department of Applied Analysis and Complex Dynamical Systems,\\
Graduate School of Informatics, \\
Kyoto University, Kyoto 606-8501, Japan\\
$^2$Center for Materials Research and Technology,\\
School of Computational Science and Information Technology,\\
and Department of Physics,\\
Florida State University, Tallahassee, Florida 32306-4350, USA
}
\date{\today }
\maketitle
\begin{abstract}
The Ginzburg-Landau model below its critical temperature
in a temporally oscillating
external field is studied both theoretically and numerically.
As the frequency or the amplitude of the external force is changed,
a nonequilibrium phase transition is observed. This transition
separates spatially uniform, symmetry-restoring oscillations from
symmetry-breaking oscillations.  Near the transition a perturbation
theory is developed, and a switching phenomenon is
found in the symmetry-broken phase.
Our results confirm the equivalence of the present
transition to that found in Monte Carlo simulations of kinetic Ising systems
in oscillating fields,
demonstrating that the nonequilibrium phase transition in both cases
belongs to the universality class of the equilibrium
Ising model in zero field. This conclusion is in agreement with
symmetry arguments [G.~Grinstein, C.~Jayaprakash, and Y.~He,
Phys.\ Rev.\ Lett.\ {\bf 55}, 2527 (1985)]
and recent numerical results [G.~Korniss, C.~J.\ White, P.~A.\ Rikvold,
and M.~A.\ Novotny, Phys.\ Rev.\ E (submitted)].
Furthermore, a theoretical result for the structure function of the local
magnetization with thermal noise, based on the Ornstein-Zernike
approximation, agrees well with numerical results in one dimension.

\end{abstract}

\section{Introduction}
\label{sec:1}

Bistable systems that are driven between their two states by a
periodically oscillating
external force are common, both in nature and technology.
A few examples are hysteresis in ferromagnetic
\cite{WARB1881,EWIN1882,JIAN95,JIAN96}
and ferroelectric \cite{ISHI71,HASH94,BEAL94} materials
driven by oscillating applied fields,
electrochemical adsorbate systems driven across a phase transition
by an oscillating electrode potential \cite{SMEL99,MITC00B,MITC00C},
and liquid crystals driven through a phase transition by pressure
oscillations \cite{CHEN96}.
In this paper we use magnetic language, henceforth referring to
the order parameter as the magnetization and the oscillating force as
the magnetic field.

When the field oscillates
at a sufficiently low frequency, the driven system essentially follows
the field, switching between its two states in a symmetry-restoring
oscillation (SRO) with the same period,
provided that the amplitude of the external force is larger than a critical
value which depends on the temperature and the system's spatial dimension.
At high driving frequencies, on the other hand, the
system is unable to follow the field and instead settles down to a
symmetry-breaking oscillation (SBO) around one or the other of its
zero-field stable states.

Over the last decade it has become evident that the boundary between
the SRO and SBO regimes corresponds to a singularity that appears to
have all the hallmarks of a genuine second-order phase transition.
It is therefore appropriate to consider the SRO and SBO as {\it dynamic
phases\/} of this far-from-equilibrium system.
Characteristic features of this nonequilibrium phase transition 
include a power-law dependence of the amplitude of the SBO
on the amount by which the frequency $\Omega$ exceeds its field and temperature
dependent critical value
$\Omega_c$, as well as critical slowing-down \cite{ACHA97D}.
The probability density of the period-averaged magnetization
exhibits a one-peak structure
for $\Omega < \Omega_c$ and a two-peak structure for $\Omega >\Omega_c$
\cite{SIDE98,SIDE99,RIKV00,KORN00,KORN00B}. In spatially extended
bistable systems,
such as the two-dimensional kinetic Ising model below its critical
temperature, the transition also displays a divergent correlation length and
finite-size scaling properties analogous to those familiar from
equilibrium phase transitions \cite{SIDE98,SIDE99,RIKV00,KORN00,KORN00B}.
It has become common to refer to this symmetry-breaking
transition as the ``dynamic phase transition'' (DPT).

The DPT was first observed in numerical solutions of a deterministic
mean-field equation of motion for a ferromagnet in an oscillating field
\cite{TOME90,MEND91}, and it has subsequently been seen and studied in
numerous Monte Carlo (MC) simulations of kinetic Ising systems
\cite{ACHA97D,SIDE98,SIDE99,RIKV00,KORN00,KORN00B,LO90,ACHA95,%
ACHA97C,ACHA98,ACHA99,BUEN00}, as well as in further mean-field studies
\cite{ACHA97D,ACHA95,ACHA98,ZIMM93A,BUEN98}.
It may also have been observed experimentally in ultrathin Co films on
Cu(100) \cite{JIAN95,JIAN96}.
Reviews of earlier research on the DPT and related phenomena
are found in Refs.~\cite{ACHA94,CHAK99}.

Finite-size scaling analysis of MC data for the DPT in the two-dimensional
kinetic Ising model at sub-critical temperatures
provides strong numerical evidence that this
nonequilibrium critical phenomenon belongs to the same universality
class as the equilibrium phase transition in the two-dimensional Ising model
in zero field \cite{SIDE98,SIDE99,RIKV00,KORN00}.
While this result may seem surprising at first,
it is consistent with a symmetry argument by Grinstein,
Jayaprakash, and He \cite{GRIN85}. This argument states that continuous
ordering transitions of fully probabilistic cellular automata with Ising-like
``up-down'' symmetry (of which the
kinetic Ising model in an oscillating field is an
example) should fall in the same universality class as the
corresponding Ising model in equilibrium.
This implies that such a cellular automaton
should possess an underlying coarse-grained effective
Hamiltonian at sufficiently large length scales, which
determines its universality class.

The purpose of the
present paper is to elucidate the origin of the DPT and to clarify the
statistical characteristics of the dynamics near the DPT, subject
to thermal noise. To this effect we consider a
time-dependent Ginzburg-Landau model with thermal
noise. The equation of motion for the noise-free version of this model is
\begin{equation}
\dot{\psi }({\bf r},t)=\psi -\psi^3+\nabla^2\psi +h\cos (\Omega t) \;,
\label{eq:1}
\end{equation}
where $\psi({\bf r},t)$ is the continuous scalar magnetization field,
and $h$ and $\Omega$ are the amplitude and frequency of the spatially
uniform external magnetic field, respectively. In zero
applied field, Eq.~(\ref{eq:1})
is identical to the conventional Ginzburg-Landau
equation for the Ising model {\it below\/} its critical temperature.
The effects of thermal noise on the system are expressed by
the stochastic differential equation
\begin{equation}
   \dot{\psi }({\bf r},t)=\psi -\psi^3+\nabla^2\psi+h\cos (\Omega t)
+R({\bf r},t) \;,
\label{eq:2}
\end{equation}
where $R({\bf r},t)$ is a Gaussian white thermal noise.
In this paper we show that Eqs.~(\ref{eq:1}) and~(\ref{eq:2}) give rise to
a bifurcation line in the ($\Omega,h$) plane. Near this bifurcation line the
equations yield an effective Hamiltonian for a {\it dynamic order
parameter\/}.
This effective Hamiltonian
is in the same universality class as the equilibrium Ising
model in zero field, and its existence provides explicit confirmation of
the symmetry argument of Ref.~\cite{GRIN85} for this far-from-equilibrium
system.

Equations (\ref{eq:1}) and~(\ref{eq:2}) with $h=0$ give rise to two
degenerate ordered solutions only for systems of spatial dimension $d \ge 2$
at temperatures below criticality. These conditions will be
assumed hereafter, unless otherwise is explicitly stated.

The present paper is organized as follows.  In Sec.~\ref{sec:2} we show
that the spatially uniform oscillation of Eq.~(\ref{eq:1}) undergoes a
bifurcation as $\Omega $
is increased, which separates the symmetry-restoring and symmetry-breaking
dynamic phases.
In Sec.~\ref{sec:3} we develop a Landau expansion near the bifurcation,
which is used to explain the switching phenomenon
observed in a system subject to thermal noise.
In Sec.~\ref{sec:3C} we show that theoretical results for spatial power spectra
of spin fluctuations (structure functions) obtained by the Landau expansion are
in good agreement with numerical experiments for a one-dimensional
system. A summary and conclusions are given in Sec.~\ref{sec:4}.

\section{Bifurcation of the symmetry-restoring oscillation}
\label{sec:2}

In this Section we concentrate on the uniform solutions of the
the noise-free system described by Eq.~(\ref{eq:1}).
The effects of spatial fluctuations and thermal noise will be
discussed in Sec.~\ref{sec:3}.

It is easy to see that Eq.~(\ref{eq:1}) has a
spatially uniform oscillation,
\begin{equation}
   \dot{\psi }(t)=\psi -\psi^3+h\cos (\Omega t) \;,
\label{eq:3}
\end{equation}
provided that spatially periodic boundary conditions are used.
Without loss of generality, $h$ and $\Omega $ are taken as positive.
Eventually, $\psi (t)$ always exhibits a periodic
oscillation of frequency $\Omega$ for any choice of $h$,
$\Omega $, and the initial value $\psi (0)$. It exhibits  no other periodic
or chaotic oscillations.  This is so because the dynamical
system (\ref{eq:3}) is dissipative and has only two degrees of freedom.

One should be careful when discussing the dynamics
near $\Omega =0$.  By shifting time as $t\rightarrow t-\pi /(2\Omega )$,
Eq.~(\ref{eq:3}) reduces to
\begin{equation}
\dot{\psi }(t)=\psi -\psi^3+h\sin (\Omega t) \;.
\label{eq:4}
\end{equation}
If one puts $\Omega =0$ in Eqs.~(\ref{eq:3}) and (\ref{eq:4})
while keeping $h$ finite,
they have different fixed points.  The period $T \;(\equiv 2\pi /\Omega )$
of the applied field tends to infinity as $\Omega \rightarrow 0$.
One should therefore
discuss the long-time behavior of $\psi $ $(t\gg T)$ at finite
$\Omega $, and then take the limit $\Omega \rightarrow 0$.
The above discrepancy originates from the interchange of the limits
$t\rightarrow \infty $ and $\Omega \rightarrow 0$.  If one takes the
limits correctly, the long-time behaviors of Eqs.~(\ref{eq:3}) and
(\ref{eq:4}) give the same results.

For $h=0$, $\psi (t) $ eventually approaches one of
the stable fixed points, $\psi_0=\pm 1$.  Then, under an applied field
$h\cos (\Omega t)$ with a small amplitude, it is easy to see that
$\psi (t)$ exhibits a periodic oscillation.  In fact, to first order
in $h$, Eq.~(\ref{eq:3}) is solved by
\begin{equation}
\psi (t)=\pm 1+\frac{h}{4+\Omega^2}\left[ 2\cos (\Omega t) +
\Omega \sin (\Omega t)\right]
\label{eq:5}
\end{equation}
for $t\rightarrow \infty $.  This is a SBO
since $\psi (t)$ oscillates near either $\psi =+1$ or
$-1$, depending on the initial condition.  We thus expect that
Eq.~(\ref{eq:3})
exhibits a symmetry-breaking periodic oscillation in the regime of
relatively weak $h$.

Let $\psi (t)$ be a solution of Eq.~(\ref{eq:3}).  It is then easy to show that
$\hat{\psi }(t)$ given by
\begin{equation}
\hat{\psi }(t) =-\psi \left(t+\frac{T}{2} \right)
\label{eq:6}
\end{equation}
is also always a solution of Eq.~(\ref{eq:3}), including even the transient
process.  As a special case (see Eq.~(\ref{eq:27}) below),
Eq.~(\ref{eq:3}) has a particular solution with the symmetry
\begin{equation}
   \psi (t)=-\psi \left( t+\frac{T}{2} \right).
\label{eq:7}
\end{equation}
If the dynamical behavior satisfies the symmetry (\ref{eq:7}), one obtains
\begin{equation}
   \int_0^T\psi (t)e^{i\ell \Omega t}dt=0\ ,\ (\ell =0,\pm 2, \pm 4,
\cdots ) \;.
\label{eq:8}
\end{equation}
However, the fact that the system (\ref{eq:3}) has the symmetry (\ref{eq:7})
does not necessarily mean that the dynamical
behavior always exhibits this symmetry.  In fact, as discussed above for
small $h$ and as shown below, Eq.~(\ref{eq:3})
has a stable symmetry-breaking solution for a certain range of
$h$ and $\Omega $.

As discussed above, the dynamics in a weak external field shows a
SBO.  This implies that the SRO,
if it exists, should do so for a relatively large amplitude of
the external field.  This also suggests that there should exist a
transition between the SBO and the SRO,
provided that the SRO stably exists.  This is
an immediate consequence of the symmetry consideration.

\begin{figure}[tb]
\vspace*{-1.0in}
\centerline{\psfig{figure=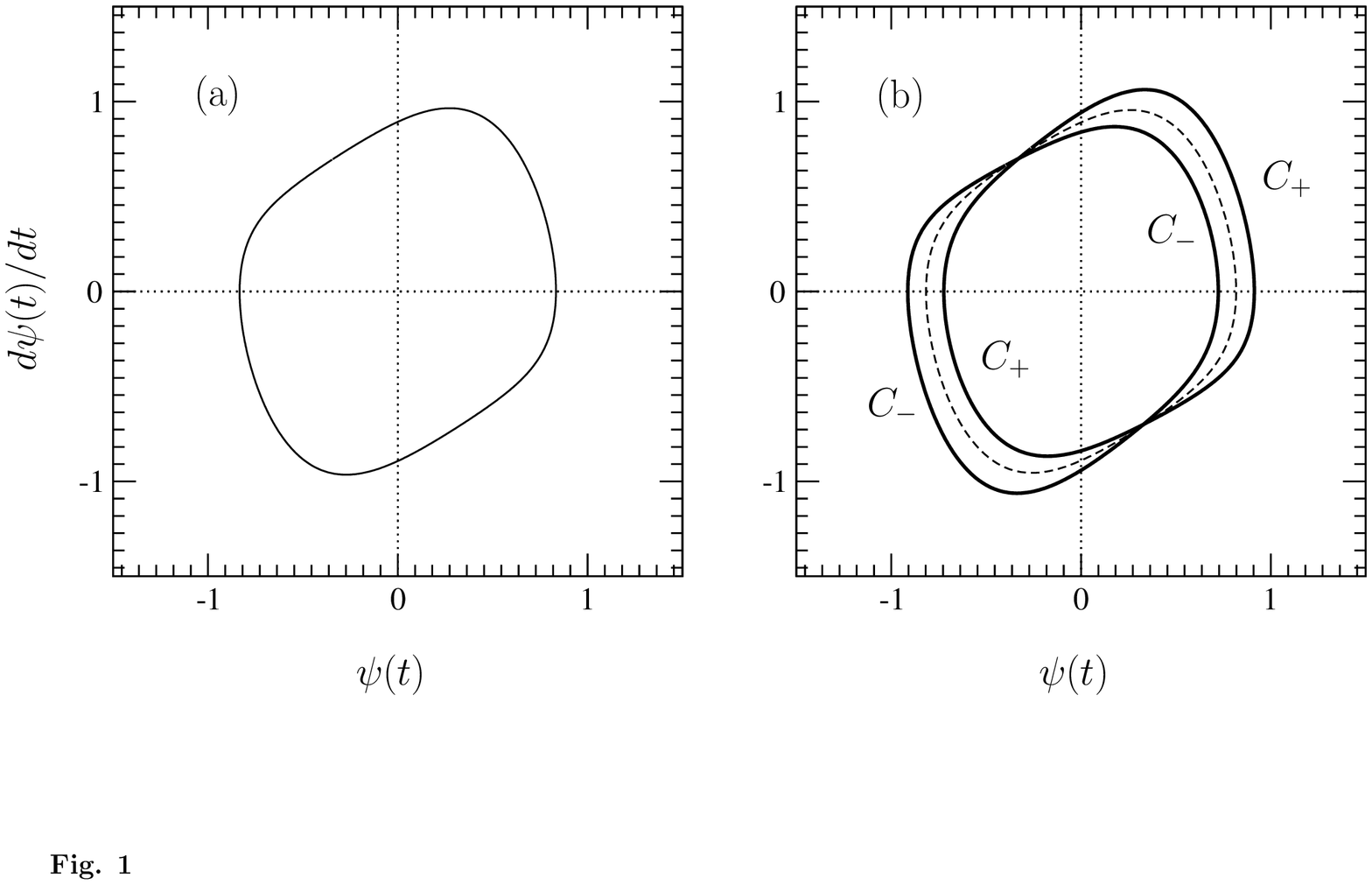,width=6.00in,angle=90}}
\caption[]{
Limit-cycle attractors for parameter values (a)
$h=1.0$, $\Omega =1.08$ and (b) $h=1.0$, $\Omega =1.1$.  The phase points move
clockwise.  In (a) ($\Omega <\Omega_c\approx 1.095$),
there stably exists only one limit
cycle, which is symmetric in the sense that Eq.~(\protect\ref{eq:6})
is satisfied.  In (b) ($\Omega >\Omega_c$), the
symmetric limit cycle denoted by the dashed curve is unstable, and there
appear two stable non-symmetric limit cycles, C$_+$ and C$_-$.
}
\label{fig:1}
\end{figure}

\begin{figure}[tb]
\centerline{\psfig{figure=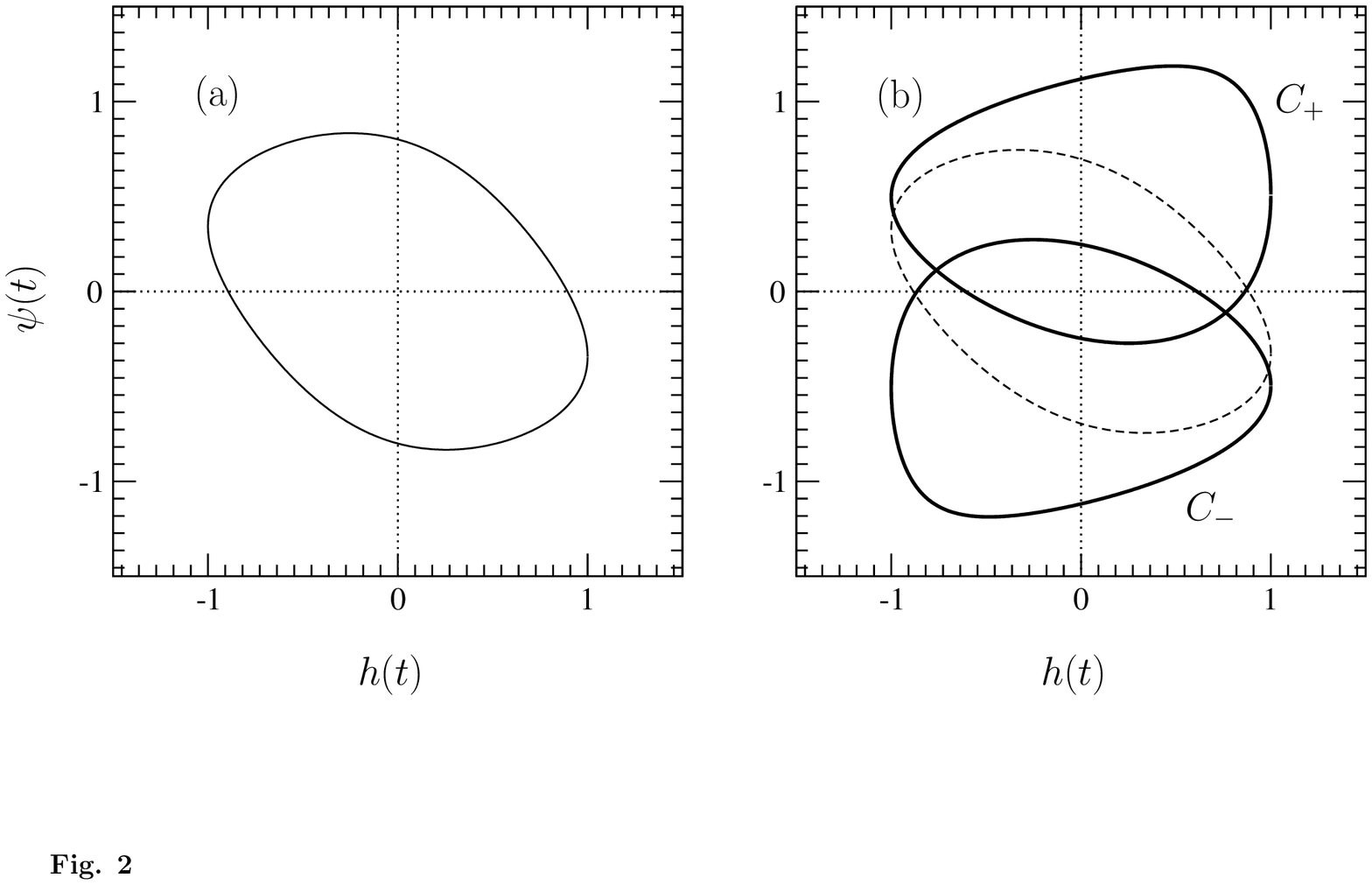,width=6.00in,angle=90}}
\caption[]{
Hysteresis loops of limit-cycle attractors, i.e., $\psi (t)$
vs $h(t)=h\cos (\Omega t)$ for (a) $h=1.0$, $\Omega =1.08$
and (b) $h=1.0$, $\Omega =1.2$.
The phase points move counterclockwise.  In (a)
($\Omega <\Omega_c\approx 1.095$), there is only one stable symmetric limit
cycle.  In (b)
($\Omega >\Omega_c$), the symmetric limit cycle (dashed curve) is unstable,
and
there exist two stable non-symmetric limit cycles, C$_+$ and C$_-$.
}
\label{fig:2}
\end{figure}

For the moment,
let us consider the parameter values $h=1.0$, and $\Omega =1.08$ and $1.1$.
For these parameter values, the system has attractors as shown
in Fig.~\ref{fig:1}.  Throughout
this paper, the numerical integration of Eq.~(\ref{eq:3})
is carried
out by using the fourth-order Runge-Kutta algorithm with the time increment
$\Delta t=T/1024$ for all frequencies.  The attractors are limit cycles
of period $T$. They have the symmetry (\ref{eq:7}) for $\Omega =1.08$,
but are asymmetric for $\Omega =1.1$.  The
above considerations suggest the existence of a phase transition between
these different characteristic oscillations.  Figure~\ref{fig:2}
shows the hysteresis
loops, i.e., the dependence of $\psi (t)$ on $h(t) \equiv h\cos (\Omega t)  $
for $\Omega $ below and above $\Omega_c $, the critical frequency separating
the interchange of the symmetric and non-symmetric oscillations.
Numerically, we find
$\Omega_c\approx 1.095$ for $h=1.0$.

\begin{figure}[tb]
\vspace*{-1.0in}
\centerline{\psfig{figure=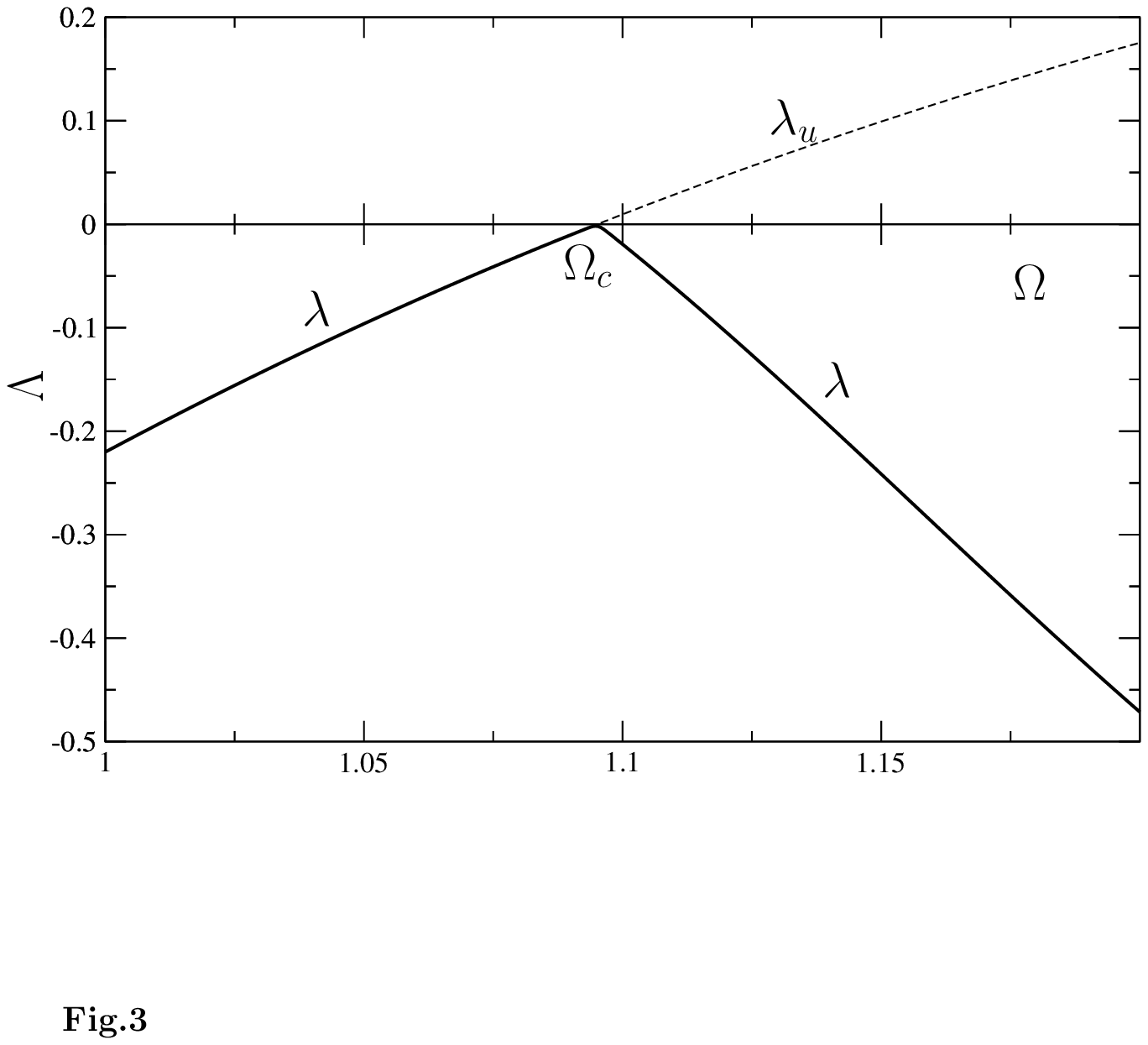,width=5.50in,angle=0}}
\vspace{-3.2in}
\caption[]{
The Floquet exponent $\Lambda$, shown vs $\Omega$ for $h=1.0$.  For $\Omega
<\Omega_c$, the limit cycle is symmetric and stable.  For $\Omega >\Omega_c$,
the dashed line is the Floquet exponent for the unstable symmetric limit
cycle,
and the solid line is the exponent for the stable symmetry-breaking
limit cycle. For details on the calculation of Floquet exponents, see the text.
}
\label{fig:3}
\end{figure}

\begin{figure}[tb]
\centerline{\psfig{figure=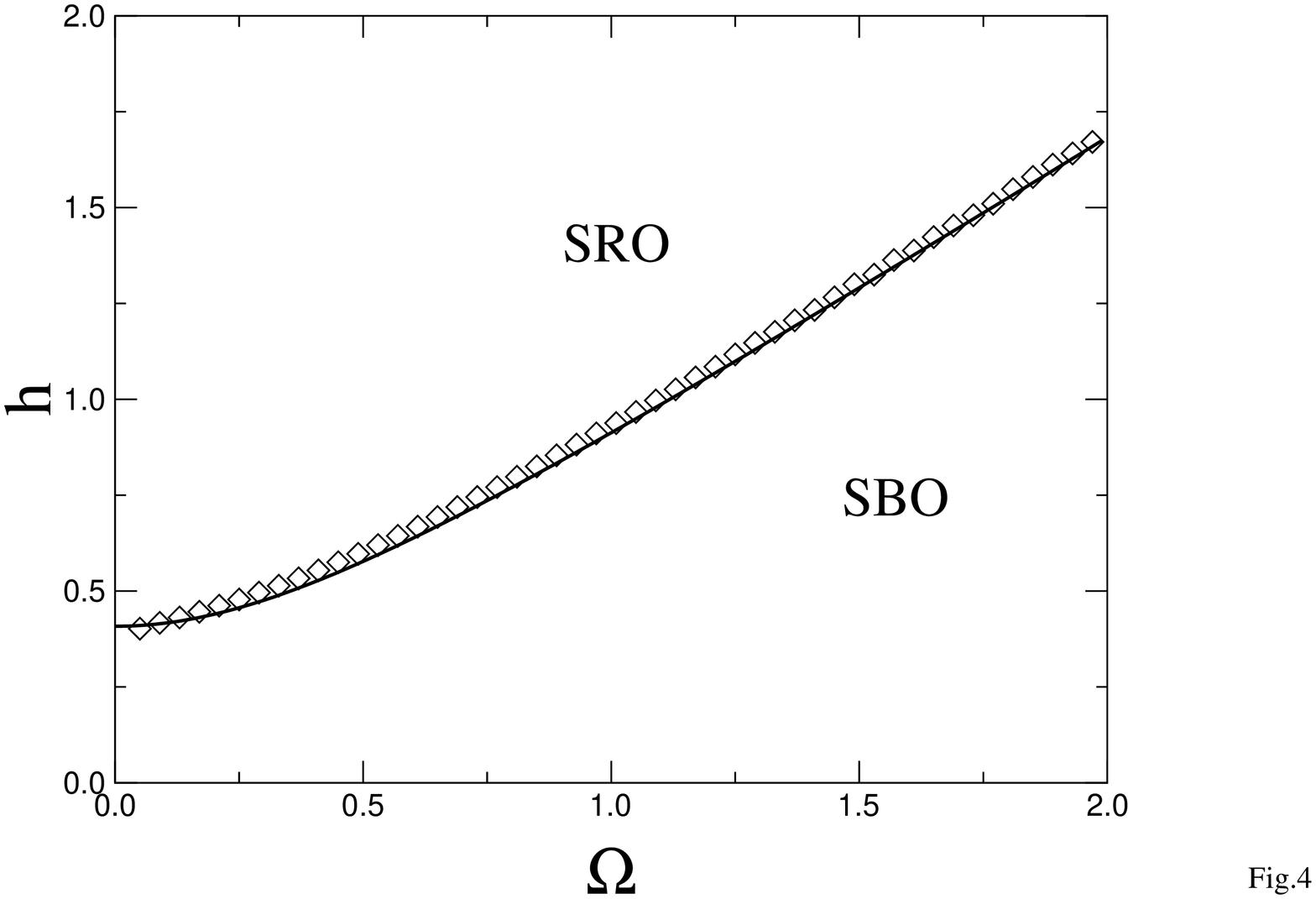,width=5.00in,angle=270}}
\caption[]{
The bifurcation curve separating the
symmetry-restoring oscillation (SRO) and the symmetry-breaking oscillation
(SBO). The numerically obtained points are represented as data points,
and the approximate
theoretical result (\protect\ref{eq:20}) as a solid curve.
}
\label{fig:4}
\end{figure}

Next we consider the stability of
the attractor with the symmetry (\ref{eq:7}), shown in
Fig.~\ref{fig:1}, as $\Omega $ is increased at fixed $h$.
The stability of a periodic oscillation is discussed as follows.  Let
$\psi (t)$ be a particular solution of Eq.~(\ref{eq:3}) on an attractor, which
may be either stable or unstable.  In order to examine its linear
stability, we seek the temporal evolution of the deviation $\delta
\psi (t)$ from this solution. Then, $\delta \psi (t)$ obeys the equation of
motion,
\begin{equation}
   \dot{\delta \psi }(t)=[ 1-3(\psi (t))^2 ]\delta \psi (t) \;.
\label{eq:9}
\end{equation}
Since $\psi (t)$ is periodic with period $T$, $\delta \psi (t)$ is
solved as
\begin{equation}
   \delta \psi (t)=B(t)e^{\Lambda t}\delta \psi (0) \;.
\label{eq:10}
\end{equation}
If we define
\begin{equation}
   \Lambda =1-3\overline{\psi^2}=1-3\frac{1}{T}\int_0^T(\psi (s))^2ds \;,
\label{eq:11}
\end{equation}
then
\begin{equation}
   B(t)=\exp \left[ -3\int_0^t\{ (\psi (t'))^2-\overline{\psi^2}  \}dt'
\right]
\label{eq:12}
\end{equation}
is a periodic function of period $T$, i.e., $B(t+T)=B(t)$.
Here we have defined the period-average of $f(t)$ 
as $\overline{f(t)}=T^{-1}\int_0^Tf(t+s)ds $.
The results (\ref{eq:10} -- \ref{eq:12}) follow from the
Floquet theorem \cite{FT}.
The quantity $\Lambda $ is called the Floquet exponent and indicates
the stability of the periodic oscillation under consideration, i.e.,
$\psi (t)$ is linearly stable (unstable) if $\Lambda <0$ $(>0)$.
Numerical results for $\Lambda $ calculated by Eq.~(\ref{eq:11}) are shown
in Fig.~\ref{fig:3}.  For $\Omega $ below
$\Omega_c$, the critical value for a given $h$, $\Lambda $ takes a negative
value, which is denoted by $\lambda $.  The limit cycle for $\Omega <\Omega_c$
is symmetric as shown in Figs.~\ref{fig:1}(a) and~\ref{fig:2}(a).
As $\Omega $ is gradually increased, the Floquet exponent approaches zero
and again takes a negative value for $\Omega >\Omega_c$.
Figures~\ref{fig:1}(b) and~\ref{fig:2}(b)
show the stable attractors (solid curves) corresponding to the limit cycles for
$\Omega > \Omega_c$.  Figure~\ref{fig:4} shows the stability regions of
the SRO and the SBO.  In the SBO region, one finds
that there exist two attractors, C$_-$ and C$_+$, one of which is chosen
depending on the initial condition.  For $\Omega >\Omega_c$, there is also
a symmetry-restoring {\it unstable\/} limit cycle, whose Floquet exponent is
denoted by $\lambda_u$ in Fig.~\ref{fig:3},
and whose trajectory is depicted by the dashed
curves in Figs.~\ref{fig:1}(b) and~\ref{fig:2}(b).
The transition at $\Omega_c$ is continuous,
as is expected from the frequency dependence of the Floquet exponent 
shown in Fig.~\ref{fig:3}.

\begin{figure}[tb]
\vspace*{-1.0in}
\centerline{\psfig{figure=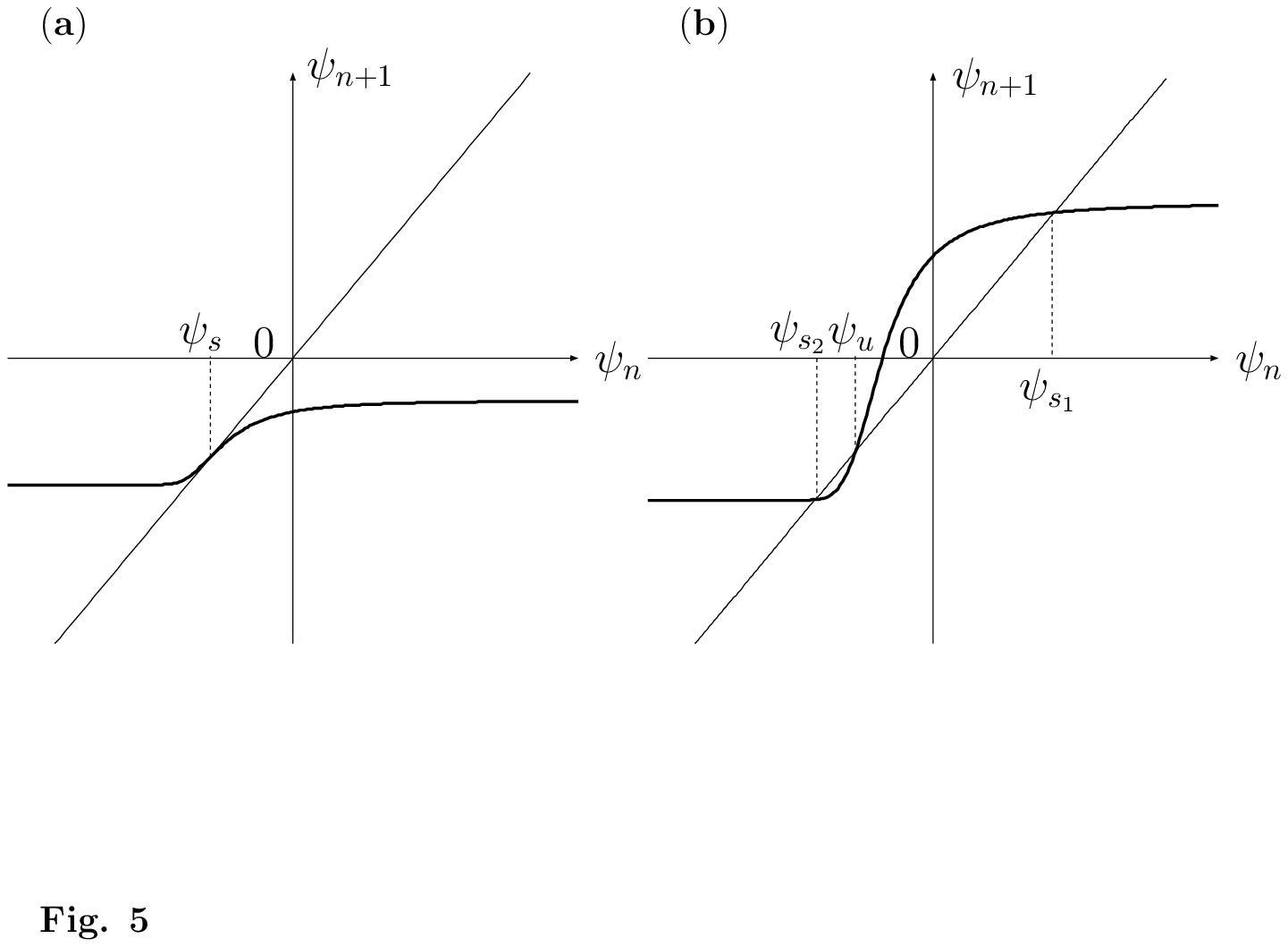,width=8.00in,angle=90}}
\vspace*{-1.0in}
\caption[]{
Stroboscopic maps $\psi_n\equiv \psi(n\cdot 2\pi
/\Omega )$ for (a) $\Omega =1.08$ $(<\Omega_c) $ and (b)
$\Omega = 1.2$ $(>\Omega_c)$ with $h=1.0$.  For
$\Omega >\Omega_c$, there appear two stable, non-symmetric limit cycles.
}
\label{fig:5}
\end{figure}

\begin{figure}[tb]
\centerline{\psfig{figure=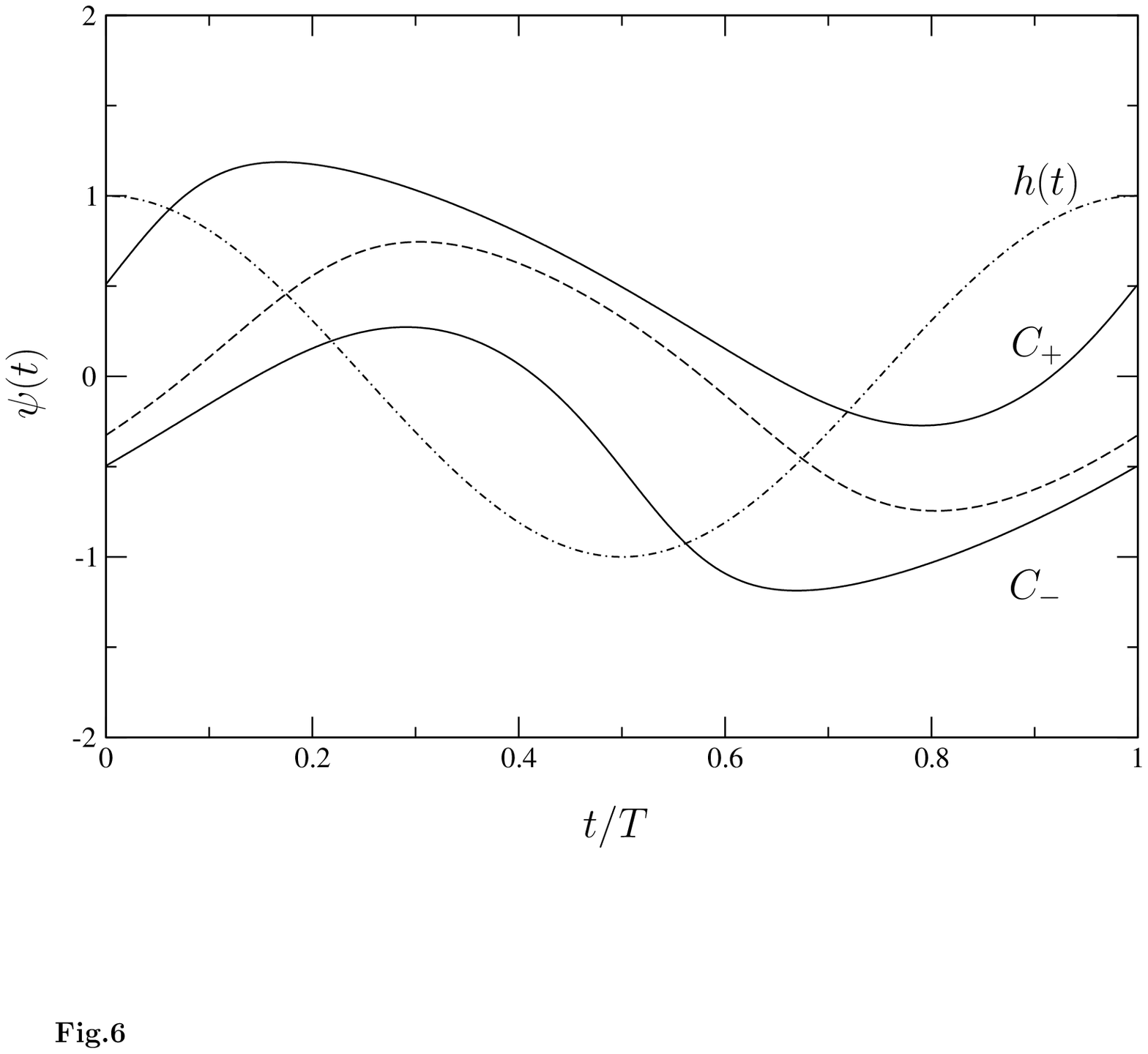,width=5.75in,angle=90}}
\caption[]{
Temporal evolutions of the stable orbits, C$_+$ and
C$_-$ (solid curves), and the unstable orbit (dashed curve)
in the SBO regime for $h=1.0$ ($< \Omega_c$) and $\Omega =1.2$ ($>\Omega_c$).
The dot-dashed curve represents $h(t)$.
}
\label{fig:6}
\end{figure}

The unstable limit cycle, i.e., the SRO for
$\Omega >\Omega_c$, is numerically obtained as follows.
Taking an initial value $\psi_n$ at time $t_n=nT$, then integrating
Eq.~(\ref{eq:3}) until $t_{n+1}=t_n+T$, we obtain $\psi_{n+1}$.
In this way we get the $\psi_{n+1}$ vs $\psi_{n}$ curve,
\begin{equation}
   \psi_{n+1}=g(\psi_n) \;.
\label{eq:13}
\end{equation}
Examples of numerically obtained $g(\psi )$ are shown in Fig.~\ref{fig:5}.
Figure~\ref{fig:5}(a) is for $\Omega <\Omega_c$, and Fig.~\ref{fig:5}(b)
is for $\Omega >\Omega_c$.  Depending on
$\Omega $, there are one or three fixed points $\psi_f$ satisfying
$\psi_f=g(\psi_f)$, which correspond to cross-sections of limit-cycle
attractors.  The stability of
a limit cycle is determined by the slope of $g(\psi )$ at $\psi =\psi_f$,
i.e., the Floquet exponent is given by
\begin{equation}
   \Lambda =\frac{1}{T}\ln |g'(\psi_f)| \;.
\label{eq:14}
\end{equation}
The unstable periodic orbit shown in Figs.~\ref{fig:1}(b)
and~\ref{fig:2}(b) is the one numerically integrated
with the initial value $\psi_u$, the unstable fixed point.  The temporal
evolutions of one unstable and two stable oscillations are shown in
Fig.~\ref{fig:6}.
One should note that if the stroboscopic map is constructed for times
$t_n=\tau +nT$, the form of $g(\psi )$ depends on $\tau $.  However, the number
of fixed points of $\psi_{n+1}=g(\psi_n )$ and the corresponding slopes,
which yield the Floquet exponents for the
fixed points, are independent of $\tau $.

The bifurcation point $\Omega_c$ depends on $h$. The theoretical
bifurcation curve given by the solid curve in Fig.~\ref{fig:4}
is determined as follows.
We first expand $\psi (t)$ as the Fourier series
\begin{equation}
   \psi (t)=\sum_{\ell =-\infty }^\infty \psi_\ell (t)e^{i\ell \Omega t} \;,
\label{eq:15}
\end{equation}
where $\psi_{-\ell }=\psi_\ell^*$.  The temporal evolution of the coefficients
$\{ \psi_\ell (t)  \} $
is assumed to be much slower than the time scale $T$.  Inserting
Eq.~(\ref{eq:15}) into Eq.~(\ref{eq:3})
and comparing the coefficients on both sides of the equation, we obtain
\begin{equation}
   \dot{\psi }_\ell +i\ell \Omega \psi_\ell =\psi_\ell
-\sum_m\sum_n\psi_m\psi_n\psi_{\ell -m-n}+\frac{1}{2}(\delta_{\ell ,1}
+\delta_{\ell ,-1})h \;.
\label{eq:16}
\end{equation}
{}From the symmetry argument, Eq.~(\ref{eq:3})
may have a solution with the symmetry (\ref{eq:7}).
If the limit cycle under consideration is symmetric, we find
from Eq.~(\ref{eq:8}) that
\begin{equation}
   \psi_\ell =0 \ \ \ {\rm for}\ \ell =0,\pm 2, \pm 4,\cdots \;.
\label{eq:17}
\end{equation}
We now consider the stability of this symmetric oscillation.
As the simplest non-trivial approximation, we use the truncation
$\ell =0$ and $\pm 1$, which yields
\begin{equation}
   \dot{\psi}_0 =\left[ 1-6|\psi_1|^2-\psi_0^2 \right] \psi_0 \;,
   \label{eq:18}
   \end{equation}
\begin{equation}
   \dot{\psi_1} +i\Omega \psi_1=\left[1-3|\psi_1|^2-3\psi_0^2 \right] \psi_1
+\frac{1}{2}h \;.
\label{eq:19}
\end{equation}
The above equations have a SRO ($\psi_0=0$) provided that $1-6|\psi_1|^2<0$.
On the other hand, for $1-6|\psi_1|^2>0$
the steady-state value of $\psi_0$ does not vanish, which implies the
emergence of a SBO.  Therefore, we find that the boundary between the
regions of stability of the
SRO and SBO is determined by $|\psi_1^{\rm ss}|^2= \frac{1}{6}$.
Combining this with the steady-state value of $\psi_1^{\rm ss}$ obtained
from Eq.~(\ref{eq:19}),
the bifurcation point $\Omega_c$ for fixed $h$ is determined by
\begin{equation}
   \Omega_c=\sqrt{\frac{3}{2} \left( h^2-\frac{1}{6} \right)}
\label{eq:21}
\end{equation}
or, equivalently,
\begin{equation}
   h=\sqrt{\frac{2}{3} \left( \frac{1}{4}+\Omega_c^2 \right)} \;.
\label{eq:20}
\end{equation}
One finds that this kind of bifurcation is observed for $h$ larger than a
critical value, $1/\sqrt{6}$ in the above approximation.
The curve given by Eq.~(\ref{eq:20}) corresponds to the transition line, which
in Fig.~\ref{fig:4} is compared with
results from numerical integration of Eq.~(\ref{eq:3}).
For $\Omega <\Omega_c$, there is only one type of periodic motion,
namely the symmetry-restoring one.
For $\Omega >\Omega_c$, on the other hand, there are
two types of oscillations: one is the unstable SRO,
and the other is represented by the two stable
SBOs, one of which is observed for a given initial condition.

Here a comment on the critical value of $h$ under a static field, $\Omega =0$,
should be added.  The above approximation yields the critical value
$1/\sqrt{6}\approx 0.408$.  On the other hand, the standard calculation
in mean-field theory yields the spinodal field
as the field where the metastable
minimum in the $\psi^4$ potential disappears.  This condition requires that
the equations,
\begin{eqnarray}
\dot{\psi } &=& \psi -\psi^3+h=0  \nonumber\\
\frac{\partial \dot{\psi }}{\partial \psi } &=& 1 -3\psi^2=0 \;,
\label{eq:22}
\end{eqnarray}
are simultaneously satisfied.  The second equation gives $\psi_{\rm spinodal }
=\pm 1/\sqrt{3}$ which, when inserted into the first equation, yields
$h_{\rm spinodal}=2\sqrt{3}/9\approx 0.385$.  This value is
about 6\% below that
obtained in the above discussion.  We carried out numerical calculations for
values of $\Omega $ as small as 0.05. The numerical results seem to
be closer to $h_{\rm spinodal}$ than to the present approximate value,
$1/\sqrt{6}$.  However, calculations at even smaller $\Omega$, which were
not feasible in the present study, would be needed to reach a firm conclusion.
A sharp decrease of the critical value of $h$ as $\Omega $ is decreased
may suggest the possibility that the transition curve may have a kind of
singularity, i.e., that $dh/d\Omega $ might diverge
as $\Omega $ approaches zero \cite{ZIMM93A}.

\begin{figure}[tb]
\vspace*{-1.0in}
\centerline{\psfig{figure=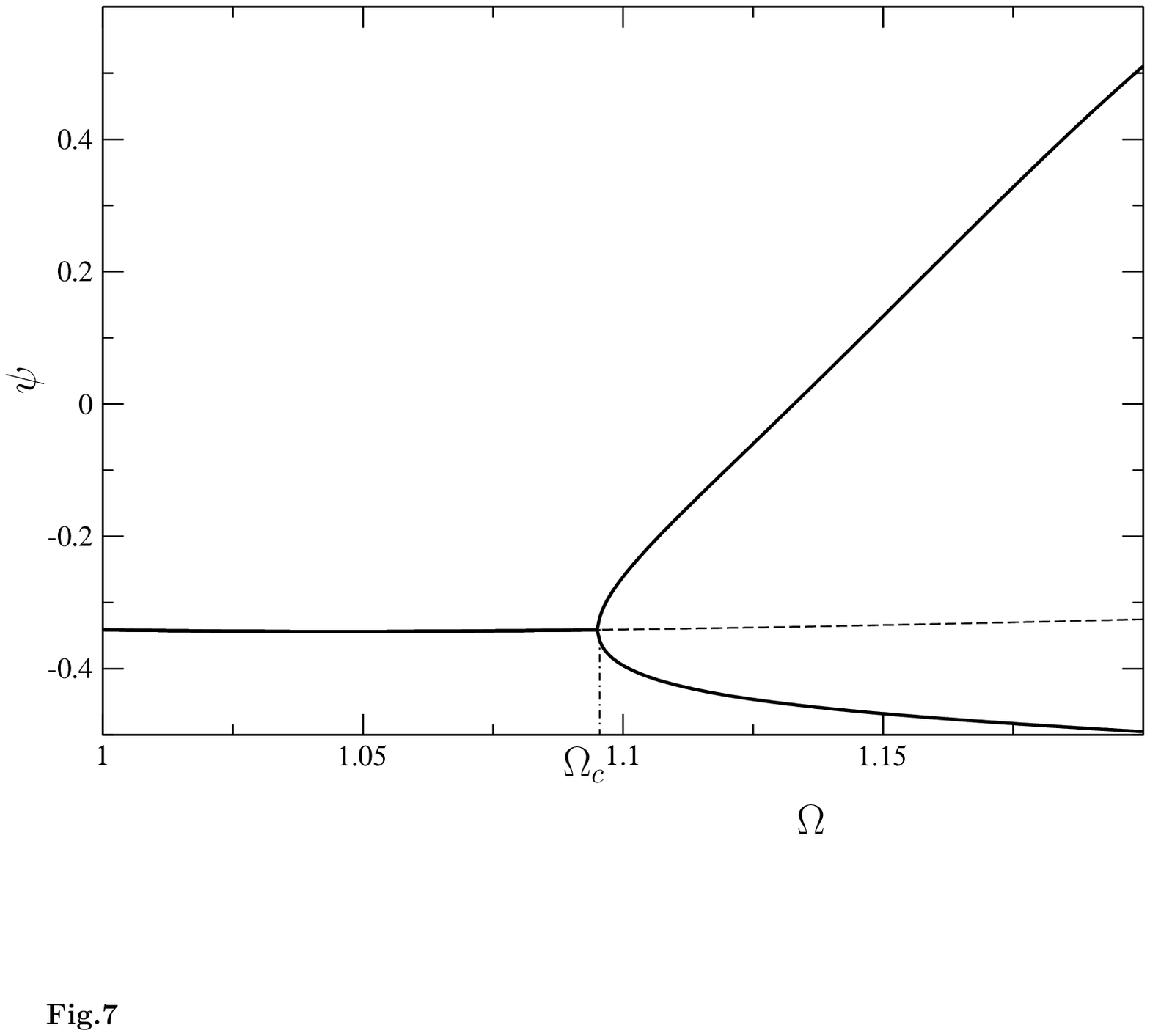,width=5.75in,angle=90}}
\caption[]{
Bifurcation diagram of $\psi_n(\equiv \psi
(n\cdot 2\pi /\Omega ))$, i.e., the fixed points of the map
(\protect\ref{eq:13}) for $h=1.0$.
{}For $\Omega <\Omega_c \approx 1.095$, there is one unique fixed point.
For $\Omega >\Omega_c$,
there exist one unstable fixed point (dashed line) and two stable fixed
points (solid curves).
}
\label{fig:7}
\end{figure}

Next we evaluate how the amplitude of the
SBO develops for $\Omega $ above $\Omega_c$.
The steady-state values $\psi_0^{\rm ss}$ and $\psi_1^{\rm ss}$ are
obtained by setting
$\psi_0^{\rm ss} = \sin \theta $ and
$\psi_1^{\rm ss}=6^{-\frac{1}{2}}\cos \theta \cdot e^{i\alpha }$.
For $\Omega <\Omega_c $, $\theta =0$, while $\theta $ is small
for $\Omega \agt \Omega_c$.  A short calculation shows that the
order parameter $\psi_0^{\rm ss}$ is asymptotically given by
\begin{equation}
   \psi_0^{\rm ss}=\pm c_1\sqrt{\Omega -\Omega_c}
\label{eq:23}
\end{equation}
with $c_1=\sqrt{8\Omega_c/(4\Omega_c^2+11)}$.  The amplitude and phase
of $\psi_1^{\rm ss}$ are given as
\begin{equation}
   |\psi_1^{ss}|=\frac{1}{\sqrt{6}}\left[ 1
-\frac{1}{2}c_1^2(\Omega -\Omega_c) \right]
\label{eq:24}
\end{equation}
and
\begin{equation}
   \alpha =\alpha_c+c_2(\Omega -\Omega_c) \;,
\label{eq:25}
\end{equation}
respectively, where
\begin{eqnarray}
\cos \alpha_c &=& \frac{-1}{ \sqrt{4 \Omega_c^2 + 1} } \;, \nonumber\\
\sin \alpha_c &=& \frac{- 2 \Omega_c}{\sqrt{4 \Omega_c^2 + 1}} \;, \nonumber\\
c_2 &=& \frac{22}{4\Omega_c^2+11} \;.
\label{eq:26}
\end{eqnarray}
The $\Omega $ dependences of the amplitudes of the SRO and SBO
are shown in Fig.~\ref{fig:7}, which was obtained from  
the stable and unstable fixed points of Eq.~(\ref{eq:13}). 

Numerical integration shows that
the time evolutions of the symmetry-breaking orbits C$_+$ and C$_-$,
which are $\psi_+(t)$ and $\psi_-(t)$, respectively,
are related as
\begin{eqnarray}
\psi_+(t) &=& -\psi_-(t+\frac{T}{2}) \nonumber\\
\psi_-(t) &=& -\psi_+(t+\frac{T}{2})
\label{eq:27}
\end{eqnarray}
(see Fig.~\ref{fig:6}), where
\begin{equation}
\dot{\psi}_\pm (t)= \psi_\pm (t)-( \psi_\pm (t))^3+ h\cos (\Omega t) \;.
\label{eq:28}
\end{equation}
The symmetry (\ref{eq:27})
is just a particular case of the general symmetry relation (\ref{eq:6}).

We find that the stable limit cycle of Eq.~(\ref{eq:3}) for $\Omega <\Omega_c$
$(\Omega >\Omega_c)$ is a symmetry-restoring (symmetry-breaking) oscillation,
and that one of the two stable SBOs is chosen,
depending on the initial condition.
Next, we examine the stability of the uniform
(stable) oscillation with negative Floquet exponent against
inhomogeneous fluctuations.
(Remember that the solution of Eq.~(\ref{eq:3}) is the uniform
solution of Eq.~(\ref{eq:1}).)
Let $\psi (t)$ be a stable solution of
Eq.~(\ref{eq:3}), which implies that its Floquet exponent is negative,
i.e., SRO for $\Omega <\Omega_c$ or SBO for $\Omega >\Omega_c$.  The
negative Floquet exponent is denoted by $\lambda $ instead of
$\Lambda $.  Next, let $\tilde{\psi }({\bf r},t)$ be the deviation from
$\psi (t)$, i.e.,
\begin{equation}
   \psi ({\bf r},t)=\psi (t)+\tilde{\psi }({\bf r},t) \; .
\label{eq:29}
\end{equation}
The Fourier transform of $\tilde{\psi }({\bf r},t)$ obeys
\begin{equation}
   \dot{\tilde{\psi }_{\bf k}}(t)=\left[ 1-3(\psi (t))^2 \right]
\tilde {\psi }_{\bf k}(t)-k^2\tilde {\psi }_{\bf k}(t) \;,
\label{eq:30}
\end{equation}
provided that the deviation is sufficiently small.  This equation is
solved as
\begin{equation}
   \tilde {\psi }_{\bf k}(t)=B(t)e^{\lambda_{\bf k}t}
\tilde {\psi }_{\bf k}(0) \;,
\label{eq:31}
\end{equation}
where $B(t)$ is again a periodic function and
\begin{equation}
   \lambda_{\bf k}=\lambda -k^2
\label{eq:32}
\end{equation}
is the linear growth rate of the Fourier mode at wave vector ${\bf k}$.
Since $\lambda <0$,
$\lambda_{\bf k}$ is always negative, which implies
that the uniform oscillation with negative Floquet exponent is linearly
stable against inhomogeneous fluctuations with any wave vector.  This
implies that the system (\ref{eq:1})
eventually approaches a spatially uniform
oscillatory motion, provided that there exists no other stable dynamical
behavior.

{}For simplicity, the values of $\Omega_c(h)$, obtained above from the 
spatially uniform solution, will be referred to as the {\it mean-field 
values\/} of $\Omega_c$. In spatially extended systems with thermal noise, 
the actual values of $\Omega_c$ are renormalized by fluctuations.

\section{Landau expansion and thermal noise effects}
\label{sec:3}

We now move on to the discussion of the spatially extended system with
local interactions, which is described by Eqs.~(\ref{eq:1}) and~(\ref{eq:2}).
The noise-free case, Eq.~(\ref{eq:1}), is discussed in Sec.~\ref{sec:3a}.
The effects of thermal noise, described by Eq.~(\ref{eq:2}), are
considered in Sec.~\ref{sec:3B}.

\subsection{Landau expansion near the bifurcation point}
\label{sec:3a}

Let $\psi_*(t)$ be the spatially uniform SRO which obeys Eq.~(\ref{eq:3})
and satisfies the symmetry (\ref{eq:7}),
and let $\Lambda $ be its Floquet exponent,
which is calculated by Eq.~(\ref{eq:11}) with $\psi (t)=\psi_*(t)$.
It is given by
$\Lambda =\lambda (<0)$ for $\Omega <\Omega_c$,
and $\Lambda = \lambda_u(>0)$ for $\Omega >\Omega_c$, in the notation used in
Fig.~\ref{fig:3}.  Expanding $\psi ({\bf r},t)$ around this SRO as
\begin{equation}
   \psi ({\bf r},t)=\psi_*(t)+B_*(t)\phi ({\bf r},t) \;,
   \label{eq:33}
   \end{equation}
where $B_*(t)$ is defined by Eq.~(\ref{eq:12}) with $\psi (t)=\psi_*(t)$,
and inserting this into Eq.~(\ref{eq:1}), we immediately find
\begin{equation}
   \dot{\phi }({\bf r},t)=(\Lambda +\nabla^2)\phi -
3\psi_*(t)B_*(t)\phi^2-(B_*(t))^2\phi^3 \;.
\label{eq:34}
\end{equation}
Note that because of the particular symmetry (\ref{eq:7}) in the
SRO phase, $B_*(t+T/2)=B_*(t)$.
Since the coefficients of the above equation are periodic in time, we may
use their time-averaged values,
noting that the characteristic time $|\Lambda |^{-1}$ of
$\phi $ near the transition is much longer than
$T$.  Making use of the symmetry relation (\ref{eq:7}), one can
prove that
\begin{equation}
   \frac{1}{T}\int_0^T\psi_*(t)B_*(t)dt
=\frac{1}{T}\int_0^T \psi_*(t)\exp \left[ -3\int_0^t\{ (\psi_*(s))^2-
\overline{\psi_*^2} \} ds  \right] dt=0 \;.
\label{eq:35}
\end{equation}
Thus, Eq.~(\ref{eq:34}) reduces to
\begin{equation}
    \dot{\phi }({\bf r},t)=(\Lambda +\nabla^2)\phi -b\phi^3
=-\tilde{\Gamma }\frac{\delta {\cal H}\{ \phi \} }{\delta \phi }
\label{eq:36}
\end{equation}
with $b=\overline{B_*^2}$ and
\begin{equation}
   \tilde{\Gamma }{\cal H}\{ \phi \}
=\int \left[ -\frac{\Lambda }{2}\phi^2+\frac{1}{2}(\nabla \phi)^2
+\frac{b}{4}\phi^4  \right] d{\bf r} \;.
\label{eq:37}
\end{equation}
Here, $\tilde{\Gamma }$ is a positive
constant, which will be determined in an appropriate way below.

\subsection{Switching phenomenon}
\label{sec:3B}

The thermal noise effects near the DPT were studied in
Refs.~\cite{SIDE98,SIDE99,RIKV00,KORN00}. In the present continuum model,
the thermal noise
$R({\bf r},t)$ is included in Eq.~(\ref{eq:2}) as a Gaussian white noise
satisfying
\begin{eqnarray}
   \langle R({\bf r},t)\rangle &=& 0 \nonumber\\
\langle R({\bf r},t)R({\bf r}',t')\rangle
&=& 2\Gamma \delta ({\bf r}-{\bf r}')\delta (t-t') \;,
\label{eq:38}
\end{eqnarray}
where $\langle \cdots \rangle $ denotes the ensemble average.
The noise strength $\Gamma $ is proportional to the temperature of the
system.  In Refs.~\cite{SIDE98,SIDE99,RIKV00,KORN00}
the thermal noise effects were studied by observing the time evolution of the
total magnetization in two-dimensional kinetic Ising
systems. A switching phenomenon between two asymmetric oscillatory
states was observed for values of $\Omega $ slightly above $\Omega_c$.
The origin of this phenomenon in the
present continuous-spin model (\ref{eq:2})
can be understood as follows. Inserting the
expansion (\ref{eq:33}) into Eq.~(\ref{eq:2}),
and approximating the coefficients by their time averages, we obtain
\begin{equation}
   \dot{\phi }({\bf r},t)=(\Lambda +\nabla^2)\phi -b\phi^3+f({\bf r},t)
=-\tilde{\Gamma }\frac{\delta {\cal H}\{ \phi \} }{\delta \phi ({\bf r},t)}
+f({\bf r},t) \;,
\label{eq:40}
\end{equation}
where $f({\bf r},t)=(B_*(t))^{-1}R({\bf r},t)$ is a Gaussian white noise
with the strength $\tilde{\Gamma }\equiv \overline{B_*^{-2}}\Gamma $.
This is also chosen as the
value of $\tilde{\Gamma }$ in Eqs.~(\ref{eq:36}) and~(\ref{eq:37}).

Equation (\ref{eq:40})
is identical to the conventional $\phi^4$ Ginzburg-Landau equation in
zero external field with a
thermal noise term. This equation belongs to the same universality class as
the Ising model \cite{GOLD92}.
This is the mechanism of the DPT and the switching phenomenon observed in
Refs.~\cite{SIDE98,SIDE99,RIKV00,KORN00}.

Equation (\ref{eq:40}) is the central result of this paper, which makes the
connection to previous work on the DPT in kinetic Ising models. In those
studies, the local dynamic order parameter has been taken as the
period-averaged magnetization,
\begin{equation}
Q_n({\bf r}) = \overline{\psi ({\bf r},t_n)}
=
\frac{1}{T} \int_{t_n}^{t_n + T} \psi({\bf r},t) dt \;,
\label{eq:DOM}
\end{equation}
where $n = {\rm int}(t/T)$. Since $\phi({\bf r},t)$ depends on $t$ only on
time scales much longer than $T$, it can be replaced by a variable
$\phi_n({\bf r})$. Thus it is easy to show that the traditional form of
the local dynamic
order parameter is simply proportional to $\phi_n({\bf r})$:
\begin{equation}
   Q_n({\bf r})=\frac{1}{T}\int_{t_n}^{t_n+T}B(s)\phi ({\bf r},s)ds
\approx \frac{\overline{B}}{T}\int_{t_n}^{t_n+T}\phi ({\bf r},s)ds
=
\overline{B} \phi_n({\bf r}) \;.
   \label{eq:56B}
   \end{equation}
The global dynamic order parameter is simply the spatial average of
$Q_n({\bf r})$. Thus, any results that are proven for $\phi({\bf r},t)$ are
also proven for the traditional dynamic order parameter, $Q_n$.

The Fokker-Planck equation corresponding to Eq.~(\ref{eq:40}) takes the form
\begin{equation}
   \frac{\partial }{\partial t}P\{ \phi ,t \}=
\tilde{\Gamma }\int \frac{\delta }{\delta \phi ({\bf r})}\left[
P^*\{ \phi \}\frac{\delta }{\delta \phi ({\bf r})} \left(
\frac{P\{ \phi ,t \} }{P^*\{ \phi \} } \right)\right] d{\bf r} \; ,
\label{eq:41}
\end{equation}
where
\begin{equation}
   P^*\{ \phi \} \propto e^{-{\cal H^*}\{ \phi \} }
\label{eq:42}
\end{equation}
is the steady-state probability density, which has a single (double) peak
structure for $\Omega <\Omega_c$ $(\Omega >\Omega_c)$.  Here ${\cal H^*}
\{ \phi \} $ is the single- or double-peaked renormalized potential function.
For $\Omega \agt \Omega_c$, the well separation in ${\cal H^*}
\{ \phi \} $ is proportional to
$(\Omega - \Omega_c)^\beta$, where $\beta$ is the magnetization exponent
for the Ising model in the appropriate spatial dimension
\cite{SIDE98,SIDE99,RIKV00,KORN00}.

We postulate that the dynamics of the total magnetization (per unit
volume, $V$),
\begin{equation}
\phi_0(t)\equiv \frac{1}{V}\int \phi ({\bf r},t)d{\bf r}
\label{eq:43}
\end{equation}
for Eq.~(\ref{eq:40}), takes approximately two values.
This implies that the dynamics can be modeled by the Langevin equation
\begin{equation}
   \dot{\phi }_0(t)=\Lambda \phi_0(t)-b(\phi_0(t))^3 + f_0(t) \;,
\label{eq:44}
\end{equation}
where $f_0(t)$ is a random force with
\begin{eqnarray}
   \langle f_0(t) \rangle  &=& 0 \nonumber\\
\langle f_0(t)f_0(t')\rangle
&=& 2\tilde{\Gamma}_0 \delta (t-t') \;.
\label{eq:45}
\end{eqnarray}
The Fokker-Planck equation is thus approximated by
\begin{equation}
   \frac{\partial }{\partial t}P(\phi_0 ,t)=
\Gamma_0 \frac{\partial }{\partial \phi_0 }\left[
P^*(\phi_0 )\frac{\partial }{\partial \phi_0 } \left(
\frac{P(\phi_0 ,t) }{P^*(\phi_0) } \right) \right] \;,
\label{eq:46}
\end{equation}
where
\begin{equation}
   P^*(\phi_0)\propto e^{-{\cal H}_0^*(\phi_0)},
\label{eq:47}
\end{equation}
where ${\cal H}_0^*(\phi_0)$ corresponds to the
critical order-parameter distribution for the Ising model
\cite{GOLD92,VK,RISKEN}. 
Except for the absence in the volume-averaged $\phi_0$ of spatial variations, 
this result is analogous to Eq.~(\ref{eq:42}) for ${\cal H}^*(\phi)$. 

\begin{figure}[tb]
\centerline{\psfig{figure=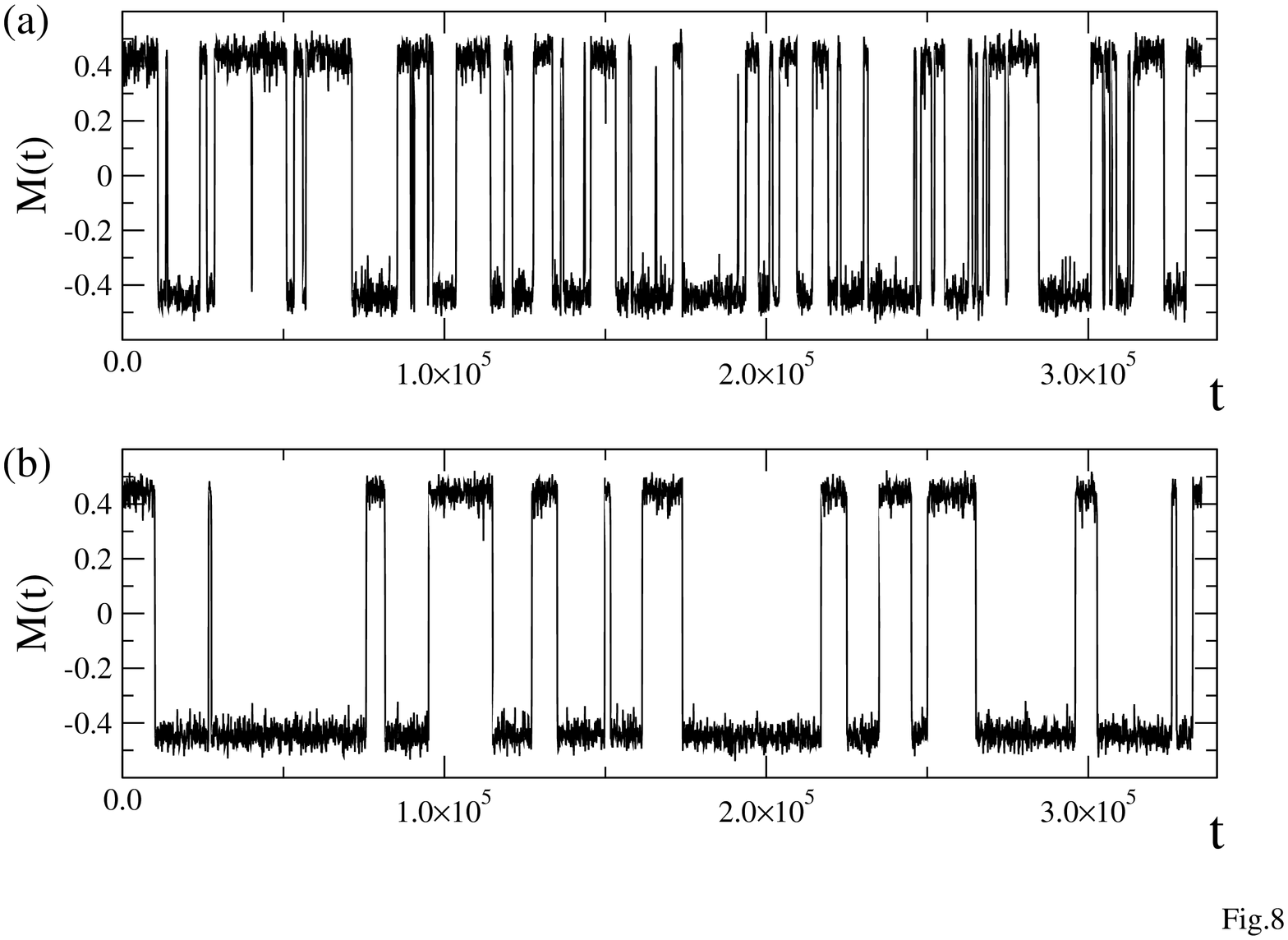,width=5.50in,angle=270}}
\caption[]{
Switching phenomenon generated by Eq.~(\protect\ref{eq:2}) in a
one-dimensional system for $\Omega $
slightly larger than the mean-field value of $\Omega_c$.
Here $M(t)=L^{-1}\int \overline{\psi (x,t)}^{(16T)}dx$ is the total
magnetization, where $L$ is the system size and
$\overline{\psi (x,t)}^{(mT)}=(mT)^{-1}\int_t^{t+mT}\psi (x,s)ds$.
Parameters are $h=1.0$, $\Omega =1.2$, ($\Omega_c \approx 1.095)$ and $\Gamma
=0.005$.
System sizes are (a) $L=16\Delta x$, (b) $L=20\Delta x$, where $\Delta x=0.5$
is the lattice spacing.  The time increment is chosen as $\Delta t=T/1024=
0.005113$.  Numerical simulations were carried out 
for $L/\Delta x=64$, 80, 96, 112, 128, 144, 160, 176, and 192. 
The average time between switching events was observed to increase 
monotonically with $L$. 
}
\label{fig:8}
\end{figure}

As an illustration of the switching behavior, 
Fig.~\ref{fig:8} shows the evolution of the total magnetization obtained by
numerically solving Eq.~(\ref{eq:2})
for $d=1$ with $\Omega $ slightly larger than 
the mean-field value of $\Omega_c$.
The numerical integration of Eq.~(\ref{eq:2}) was carried out by 
the second-order stochastic Runge-Kutta (SRK) algorithm \cite{BH98},
dividing the space into lattice points with lattice spacing $\Delta x$
(set to 0.5 throughout this paper) and using
\begin{equation}
(\nabla^2\psi )_j=\frac{\psi_{j-1}-2\psi_j+\psi_{j+1}}{(\Delta x)^2}
\label{eq:39}
\end{equation}
at the lattice site $j$ with periodic boundary conditions.
A switching phenomenon is clearly observed.

The switching phenomenon can be formulated in a different way as follows.
The temporal evolutions $\psi_+(t)$ and $\psi_-(t)$ of the symmetry-breaking
orbits, C$_+$ and C$_-$, respectively, obey Eq.~(\ref{eq:28}).
We study the additive noise effect on the dynamics, adding
a weak noise $R$, as in Eq.~(\ref{eq:2}).
We also include the spatial variation of the dynamical
variable, adding $\nabla^2\psi $.
Consider a local bistable variable.  For $\Omega >\Omega_c$, depending on the
initial condition, either C$_+$ or C$_-$ is selected, provided the
noise is absent.  If the noise is sufficiently weak, the phase point is almost
always on either C$_+$ and C$_-$.  When $\psi_+$ and $\psi_-$ are close, the
phase point can switch to the other orbit through the noise effect.  The above
picture can be mathematically formulated as follows.  Let $a(t)$ be a
variable which takes the values $+1$ $(-1)$, provided the phase point
is on C$_+$ (C$_-$)
at time $t$.  Then the temporal evolution of $a(t)$ is approximately
described by the switching dynamics.
The above picture is generalized by introducing the position dependent
variable $a({\bf r},t)$, defined by
\begin{equation}
\psi ({\bf r},t)=\frac{1-a({\bf r},t)}{2}\psi_+(t+t_0^+)+
\frac{1+a({\bf r},t)}{2}\psi_-(t+t_0^-) \;,
\label{eq:49}
\end{equation}
where $t_0^\pm $ are certain initial times. The variable $a({\bf r},t)$
indicates whether the local magnetization $\psi ({\bf r},t)$ is close to
$\psi_+(t+t_0^+)$ or $\psi_-(t+t_0^-)$.  Namely, if $a({\bf r},t)$ is near
$+1$ or $-1$, $\psi ({\bf r},t)$ is close to $\psi_+(t+t_0^+)$ or
$\psi_-(t+t_0^-)$, respectively.  Without loss of generality,
$t_0^\pm $ are chosen such that
$\cos (\Omega t_0^\pm )=1$, and therefore we put
$t_0^\pm =0$.  Inserting Eq.~(\ref{eq:49})
into Eq.~(\ref{eq:2}), after some algebra we rigorously get
\begin{equation}
\dot{a}({\bf r},t)=\frac{1}{4}(1-a^2)\left[ \left( \psi_+(t)-\psi_-(t)
   \right)^2a-3\{ (\psi_+(t))^2-(\psi_-(t))^2  \} \right] +\nabla^2a+
   g({\bf r},t)
\label{eq:50}
\end{equation}
with
\begin{equation}
g({\bf r},t)=\frac{-2R({\bf r},t)}{\psi_+(t)-\psi_-(t)} \; .
\label{eq:51}
\end{equation}
Furthermore, as long as the thermal noise is weak, the average
switching time between $a=+1$ and $-1$ is long. The
temporally periodic coefficients can therefore be replaced by their
average values, which reduces Eq.~(\ref{eq:50}) to
\begin{equation}
\dot{a}=\mu (1-a^2)a+\nabla^2a+g({\bf r},t)
\label{eq:52}
\end{equation}
with
\begin{equation}
\mu =\frac{1}{4}\overline{\left( \psi_+(t)-\psi_-(t) \right)^2}
=\frac{1}{4}\overline{(B(t))^2\left( \phi_+(t)-\phi_-(t) \right)^2} \;(>0) \; ,
\label{eq:53}
\end{equation}
where $\overline{\psi_+^2}=\overline{\psi_-^2}$ by symmetry.
The quantities $B(t)$ and $\phi_\pm (t)$ are the same as in Sec.~\ref{sec:2}.
Equation (\ref{eq:52})
again takes the form of the Ginzburg-Landau equation with
a double-well potential, with stable fixed points $a=\pm 1$,
provided that the spatial variation
of $a$ and the noise are neglected.  Equation (\ref{eq:52})
shows the switching phenomenon.

\section{Structure function in the SRO phase}
\label{sec:3C}

\begin{figure}[tb]
\centerline{\psfig{figure=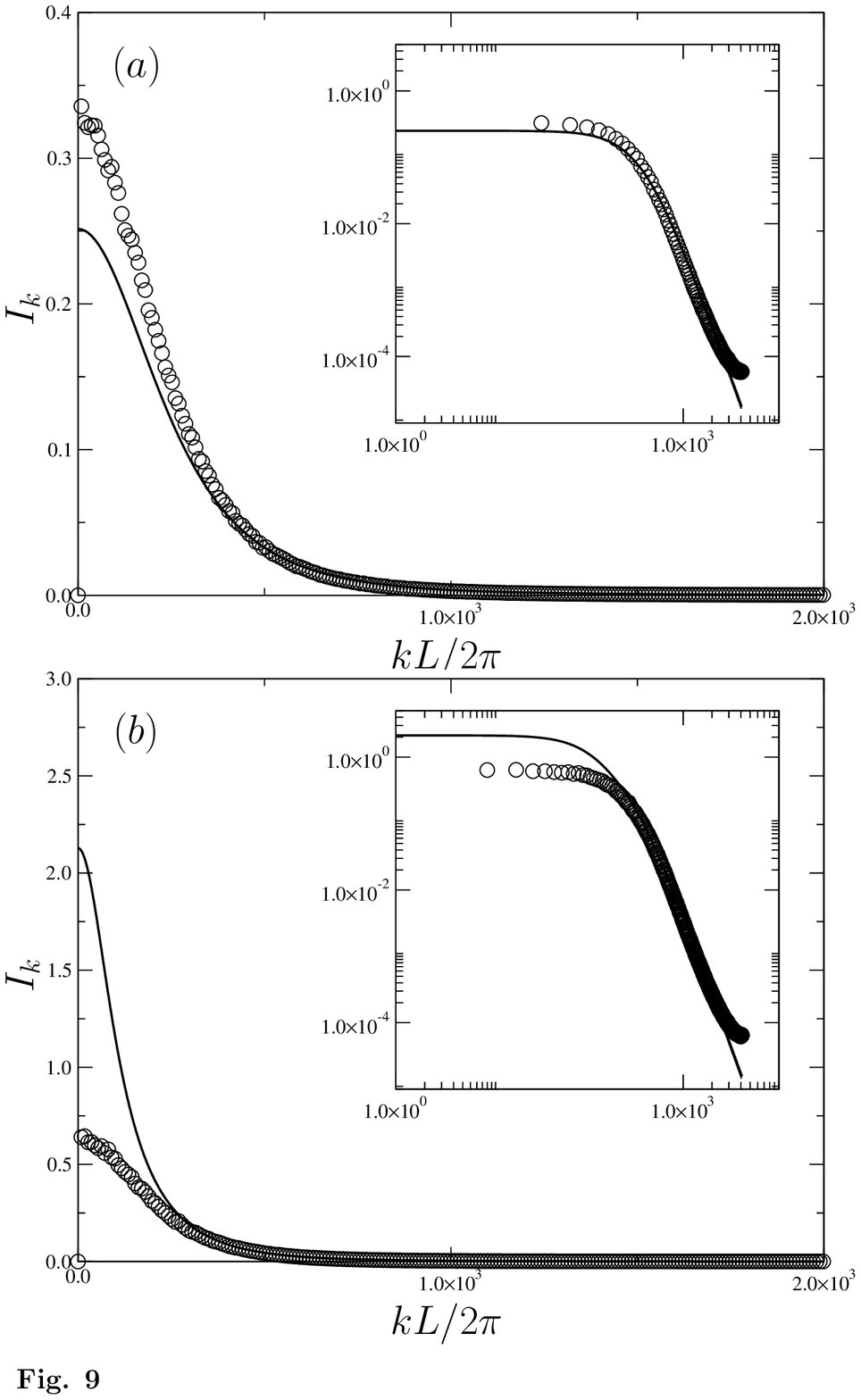,width=5.75in,angle=0}}
\caption[]{
The structure functions
for the Fourier components of $\overline{\psi  ({\bf r},t)}$
in a one-dimensional system with thermal noise in the SRO regime.
The numerically obtained results are shown as $\circ$,
while the theoretical result
of the linearized model (\protect\ref{eq:58}) is shown as a solid curve.
The insets are log-log plots.
The parameters are (a) system size $L=8192\times \Delta x$ with lattice
spacing $\Delta x=0.5$, $h=1.0$, $\Omega =1.0$, $(\lambda =-0.220)$,
and $\Gamma =0.05$,
and (b) $L, \Delta x$, $h$, and $\Gamma $ the same as in (a), while
$\Omega =1.08$, $(\lambda =-0.0307)$.
}
\label{fig:9}
\end{figure}

In this Section, we study the structure function in the SRO phase
($\lambda <0$) in a one-dimensional system. We define the structure function,
\begin{equation}
   I_{\bf k}=\langle  |\overline{\psi_{\bf k}(t)   }|^2\rangle \;,
   \label{eq:54}
   \end{equation}
for the Fourier transform $\overline{\psi_{\bf k}(t) }$ of
the period-averaged order parameter,
\begin{equation}
   \overline{\psi ({\bf r},t)}\equiv \frac{1}{T}\int_t^{t+T}
\psi ({\bf r},s)ds \;.
   \label{eq:55}
   \end{equation}
We obtain $I_{\bf k}$ by numerically solving Eq.~(\ref{eq:2})
in a one-dimensional system for several values of
$\Omega$ in the SRO phase as shown in Fig.~\ref{fig:9}.
One characteristic feature
of the numerically obtained structure function is the
$k^{-4}$ behavior observed for relatively large $k$.
Here, $I_{\bf k}$ can be
evaluated using the Landau expansion with thermal noise as follows.  By
inserting Eq.~(\ref{eq:33}) into Eq.~(\ref{eq:55}) and using the approximation 
[see also Eqs.~(\ref{eq:DOM}) and~(\ref{eq:56B})]
\begin{equation}
   \overline{\psi ({\bf r},t)}=\frac{1}{T}\int_t^{t+T}B(s)\phi ({\bf r},s)ds
\approx \frac{\overline{B}}{T}\int_t^{t+T}\phi ({\bf r},s)ds \;,
   \label{eq:56}
   \end{equation}
the structure function is evaluated as
\begin{equation}
   I_{\bf k}=\frac{\overline{B}^2}{T^2}\int_0^Tds \int_0^Tds'
\langle  \phi_{\bf k}(s)\phi_{\bf k}^*(s')\rangle \;.
   \label{eq:57}
   \end{equation}
Using the linearized form of the equation of motion for $\phi ({\bf r},t)$,
Eq.~(\ref{eq:40}), which yields 
\begin{equation}
   \dot{\phi }({\bf r},t)=(\lambda +\nabla^2)\phi ({\bf r},t)+f({\bf r},t)
   \;,
   \label{eq:58}
   \end{equation}
we immediately get the correlation function
\begin{equation}
       \langle  \phi_{\bf k}(s)\phi_{\bf k}^*(s')\rangle
=\frac{\tilde{\Gamma }}{\gamma_{\bf k}}e^{-\gamma_{\bf k}|s-s'|} \;,
   \label{eq:59}
   \end{equation}
where
\begin{equation}
    \gamma_{\bf k}=|\lambda |+k^2
   \label{eq:60}
   \end{equation}
is the damping rate for the fluctuation $\phi_{\bf k}(t)$.  Substitution
of Eq.~(\ref{eq:59}) into Eq.~(\ref{eq:57}) yields
\begin{equation}
     I_{\bf k}=p \Gamma T
G(\gamma_{\bf k}T) \;,
   \label{eq:61}
   \end{equation}
where $ p =\overline{B}^2\overline{B^{-2}} $ and
\begin{equation}
     G(x)=\frac{2}{x^2} \left( 1-\frac{1-e^{-x}}{x} \right)
   \label{eq:62}
   \end{equation}
is a scaling function.  We thus find that the structure function
$I_{\bf k}$ behaves asymptotically as $I_{\bf k} \sim \gamma_{\bf k}^{-1}$
for $\gamma_{\bf k}T\ll 1$ and as $\sim \gamma_{\bf k}^{-2}T^{-1}$
for $\gamma_{\bf k}T\gg 1$.  In the former case, the relation
$I_k\propto \gamma_k^{-1}$ is identical to the Ornstein-Uhlenbeck form which
holds for $T\rightarrow 0$.  The latter characteristic explains the $k^{-4}$
behavior in the large-$k$ regime.  This unexpected behavior originates
from the temporal averaging procedure (\ref{eq:55}), which ensures that the
interfaces between regions of positive and negative values of
$\overline{\psi_{\bf k}(t) }$ are {\it not\/} thin. As a result, $I_{\bf k}$
does not obey ``Porod's law,'' $I_{\rm k} \sim k^{-(d+1)}$ \cite{PORO},
which would yield $I_{\rm k} \sim k^{-2}$ for the present case of $d=1$.

The solid curves in Fig.~\ref{fig:9} correspond to
the theoretical result (\ref{eq:61}--\ref{eq:62}).
Although this linear model for the fluctuations agrees
quite well with the numerical simulations, particularly in the
large-wavenumber regime
and for strong thermal noise, nonlinear fluctuations play a significant
role and the linear model eventually breaks down.

\section{Summary and Conclusion}
\label{sec:4}

In this paper we used a time-dependent Ginzburg Landau model
in a temporally oscillating external field to
understand the dynamic phase transition (DPT)
observed in Monte Carlo simulations of
the corresponding kinetic Ising model below its critical temperature
\cite{ACHA97D,SIDE98,SIDE99,RIKV00,KORN00,KORN00B,ACHA95,ACHA97C,ACHA98,%
ACHA99,ACHA94,CHAK99}.
Analyzing the stability of spatially uniform oscillations, we found a
bifurcation of the symmetry-restoring oscillation (SRO),
which leads to the onset of
a symmetry-breaking oscillation (SBO). Developing a Landau expansion near the
bifurcation point, including additive thermal noise,
we obtained an effective Ginzburg Landau
Hamiltonian for the amplitude of the SBO,
which is proportional to the dynamic order
parameter which characterizes the DPT. This effective Hamiltonian has the
same form as the standard $\phi^4$ Ginzburg Landau Hamiltonian in {\it zero\/}
external field, which describes the long-range properties of the Ising model
in the critical region \cite{GOLD92}.
This result implies that the DPT belongs to the
same universality class as the equilibrium Ising model in zero external field,
in agreement with recent high-precision
numerical results from Monte Carlo simulations
\cite{SIDE98,SIDE99,RIKV00,KORN00}, as well as with a symmetry argument
which states that the equilibrium Ising universality class should encompass
all stochastic cellular automata with Ising ``up-down'' symmetry \cite{GRIN85}.

To the best of our knowledge, the work presented here is the first
in which an effective Hamiltonian has been explicitly derived for a
far-from equilibrium phase transition, confirming that the transition
belongs to the
same universality class as the equilibrium Ising model in zero field.
The result represents a significant expansion of the realm of validity of
symmetry arguments from equilibrium to nonequilibrium phase transitions.

\section*{Acknowledgments}

We appreciate useful comments on the manuscript by G.~Brown.

P.~A.~R.\ acknowledges the hospitality in 1998 of the Department of
Fundamental
Sciences, Faculty of Integrated Human Studies, Kyoto University, which
enabled the authors to begin their collaboration.

This study was partially supported by Grant-in-Aid for Scientific
Research No. 11837009 from the Ministry of Education, Science,
Sports and Culture of Japan.
Work at Florida State University was supported in part by U.S.\ National
Science Foundation grants No.~DMR-9634873 and DMR-9981815,
and by Florida State University
through the Center for Materials Research and Technology and the School of
Computational Science and Information Technology.


\newpage

\end{document}